\documentclass{sig-alternate}
\usepackage{makeidx}  
\usepackage{graphics} 
\usepackage{url}
\usepackage[subtle]{savetrees}
\begin{document}
\title{Do we have the time for IRM?:\\ Service denial attacks and SDN-based defences}
%
%
\author{Ryan Shah and Shishir Nagaraja\\
  University of Strathclyde\\
  \tt{\{ryan.shah,shishir.nagaraja\}@strath.ac.uk}}

\maketitle              

\pagenumbering{gobble}

\begin{abstract}
Distributed sensor networks such as IoT deployments generate large quantities of measurement data. Often, the analytics that runs on this data is available as a web service which can be purchased for a fee. A major concern in the analytics ecosystem is ensuring the security of the data. Often, companies offer Information Rights Management (IRM) as a solution to the problem of managing usage and access rights of the data that transits administrative boundaries. IRM enables individuals and corporations to create restricted IoT data, which can have its flow from organisation to individual control -- disabling copying, forwarding, and allowing timed expiry. We describe our investigations into this functionality and uncover a weak-spot in the architecture -- its dependence upon the accurate global availability of \emph{time}. We present an amplified denial-of-service attack which attacks time synchronisation and could prevent all the users in an organisation from reading any sort of restricted data until their software has been re-installed and re-configured. We argue that IRM systems built on current technology will be too fragile for businesses to risk widespread use. We also present defences that leverage the capabilities of Software-Defined Networks to apply a simple filter-based approach to detect and isolate attack traffic.
\end{abstract}

%
\section{Introduction}
\label{intro}

Led by the intense desire to sense ubiquitously, measure universally,
and apply data analytics to sensed information in the hope of adding
value, governments, industry, society, and the individual are hastily
adopting the vision of the Internet-of-everything. In the majority of
cases the primary aim is to collect data, apply analytics and sell the
intelligence gathered further on within the ecosystem. In the case of
industrial IoT systems, organisations hope to adopt a data-driven
approach towards managing, assessing and verifying their business
workflows. Manufacturing industries are keen to understand how coarse
to very-fine grained measurements about their processes can add value
to their bottom line; service industries are similarly interested in
obtaining fine-grained measurements with the view of moving from a
periodic maintenance cycle to a predictive maintenance cycle -- i.e
by using automated and sensed data-driven approaches, firms believe
they can predict when components within a system are showing signs of
failure. For instance, a conventional lift requires periodic
inspection and replacement of parts which are most worthy of
operation but that must be replaced in order to achieve a working lift
with high probability until the next maintenance cycle (i.e. high
availability). However, such conventional models of periodic
maintenance schedules, tend to overestimate component failures
resulting in excess maintenance expenditure and missing failures
between maintenance cycles. On the other hand, a data-driven approach
where the lift is equipped with speed, weight, cable tension, and
shaft sensors, engenders predictive failure models that promise better
results --- machine learning techniques applied to the data streams
promise to detect problems as they unfold and prior to catastrophic
component failure. Thus maintenance cycles can be sparser thus saving
money for the maintenance company.

Nice as these ideas are in theory there are some fundamental
challenges with data management that arise in IoT environments. To
generate usable intelligence from hyperconnected networks of sensors,
the defender (network operator) must collect the information at a
centralised location in order to run machine learning algorithms over
their data. A cloud storage option is a natural choice for such a
storage location. However, it isn't always possible to run all
analytics in a single datacentre. In order to take advantage of
specialist analytics services, typical workflows require IoT data to
be sent across administrative boundaries. So how does an organisation
secure data that is stored beyond the customer's datacentre.

Microsoft provides a solution to the problem of managing information
rights over data that transfers across administrative boundaries such
as data centres, cloud providers, and managed analytics
services. Microsoft proposes to accomplish this using Azure Rights
Management (RMS) technology~\cite{grothe2016break}. RMS can be used to
manage IoT data from any network or IoT hub as long as the data is
stored on an Azure cloud.

RMS hopes to make data behave like physical objects. RMS ensures that
the access control metadata placed on IoT data is enforced on remote
files, i.e. even if the data is moved from its location on the cloud
on to a specialist workstation, a different cloud or data centre, or
copied to a data storage that's not under the control of an IT
organisation. RMS also promises audit and monitoring support to stored
data, for instance, the IoT network operator is notified when a
remotely located datafile is accessed, processed, or moved to another
location where this information can be accessed by someone else.

While much attention within IoT security has been devoted to the
analysis of security and privacy protocols for inter-device and
device-hub communication~\cite{hossain2015towards}, much less
attention has been paid to Rights Management services used by
organisations to manage the huge amounts of data that IoT deployments
are expected to generate. Naturally, the security of these rights
management services is of crucial importance. If the rights management
can be compromised, then at best the IoT network operator will lose
control of their data. Worse still, an attacker might deny access to
the IoT network operator itself, resulting in service-denial attacks
on the IoT infrastructure. This can have serious consequences beyond
the mere loss of data.

IoT based telemetry and monitoring drive modern safety regimes where
predictive analytics drives (reduced) maintenance frequency and
replace traditional maintenance cycles.  Without measurement data, the
safety of the appliance or system is at risk. Therefore, a denial of
service attack on IoT data can directly lead to a safety
compromise. In other words, a security attack translates into a safety
problem. In safety-critical applications, such as elevators, medical
device operation, or drug manufacturing, the absence of maintenance
information might require the appliance to be shut down for safety
reasons, with the possibility of cascading failures further down the
dependency chain.

With RMS, an attacker won't need to find individual device
vulnerabilities to exploit. Instead, if the network operators rights
to their data can somehow be revoked or suspended, then service denial
attacks again RMS can be generalised across IoT networks without
attackers being required to know, understand or exploit the devices
and their vulnerabilities within a target network. In this paper, we
propose service denial attacks on an IoT deployment using the Azure
Rights Management Service.


\section{Information Rights Management}
\label{irm}

We introduce Digital Rights Management (DRM) as a precursor to Information Rights Management, which is derived from DRM, and discuss the successes and limitations of both types of rights management.

DRM is a collection of technologies developed for restricting the use
of hardware and copyrighted work. This includes the control of access,
modification and distribution of content, and the systems that enforce
these policies~\cite{rosenblatt2001digital}. It has allowed for the
prevention of unauthorised distribution of copyrighted digital media,
and DRM technologies have been used in a variety of technologies
including, but not limited to: documents, film and music. DRM has also
been used in conjunction with other technologies, such as using
steganography to provide DRM control information over insecure
communication channels~\cite{van2002steganographic}. DRM can be
generalised into four components - digital rights to manage,
encryption, license management and a DRM-enabled
client~\cite{ku2004survey}. A packaging server is used to distribute
license data, domain certificates and packaged content to other
servers which manage the different aspects of a DRM system. License
data is sent to a license server which is used to request and issue
licenses to DRM-enabled clients. Domain certificates are registered
with and issued by a domain controller to the clients, and finally a
distribution server (or multiple) request and deliver the protected
content to the clients.

Typically, DRM uses encryption which is applied to the content to be
protected, such that it can only be ``unlocked'' with the correct
key. Key exchange is a critical part of DRM and requires a
\textit{root of trust}~\cite{armcitation}. This involves the
distributor only providing the keys to software, services or devices
it trusts. The distribution of the keys can vary on many factors
including the device, the content and security levels required. It can
range from merely providing keys to devices which share a correct
token, to secret keys embedded in the device used for pairing with
advanced levels of encryption. The primary advantage of DRM is that
its existence provides a level of assurance to the owner when
allowing the electronic distribution of their content.


IRM is derived from DRM and involves technologies that protect
sensitive information from unauthorised access. It started as a
feature that allowed users to control the flow of email and office
documents such as word-processed files or spreadsheets and was
expanded to include many other types of data after 2010. While DRM is
primarily used to protect intellectual property from patent
infringement and piracy, IRM focuses on protecting sensitive data ---
especially data that is exchanged with external entities outside of
its originating organisation. The main difference between IRM and DRM
is that a true IRM system separates the information from its control
such that either can be accessed, manipulated and distributed
separately~\cite{bartlett2009information}.

IRM encrypts data and applies IRM rules which enforce access policies
to allow or deny specific activities, such as the data being read-only
or blocking any data from being copied from a document. A client who
is entitled to access the sensitive data must first be registered with
the IRM server. After the user is authenticated, the server will
download some code to the client device. Every time the user requests
a new document from the server or accesses the existing document(s) on
their device, the code on the device must reauthenticate on the IRM
server. This allows for a key to be downloaded which in turn, decrypts
the document and determines the access policies the client is entitled
to. Some IRM services allow for time-limited offline access
privileges, for example to those who want to peruse documents in
places with limited access to the Internet. The benefit to IRM is that
these enforced policies persist even when data is sent externally,
such that IRM sealed documents can remain secure no matter where they
are accessed. With this said, IRM solutions require a client user to
have specialised IRM software installed on their device in order to
access data with IRM protection. For this reason, many organisations
limit the use of IRM protection to data that requires it.

\paragraph{Rights management and IoT:} Bauer et al. stress the importance of
data provenance throughout the life-cycles of IoT devices to create a
trusted and secure IoT environment~\cite{bauer2013data}. They suggest
the use of DRM technologies to control and validate sensitive meta
information such as provenance data. Matthieu and Ramleth propose an
alternate the use of IRM for managing security and access rights
within an IoT context~\cite{matthieu2015security}. They propose a
cross-domain messaging architecture for IoT devices, where a remote
analytics service performs computations over data supplied by IoT
deployments spanning administrative and organisation boundaries,
whilst restricting the use of the supplied data for the stated
purpose. Huckle et al.~\cite{huckle2016internet} identify how IoT and
blockchain technologies can use digital rights management to enable
applications that enforce usage rights in the physical world, such as
managing access rights to resources among tenants in a shared home or
within a holiday home.

\subsection{Limitations}
A disadvantage of DRM is that it does not guarantee enforcement in any
fundamental way -- once an approve device has access to the data, a
large number of side-channel attacks can be mounted. Further, the
restrictions can result in a significant insult rate as the rights
management system tries to distinguish between legitimate and
illegitimate users~\cite{armcitation}. While consumers may feel
insulted at treated as a potential \textit{pirate user}, in a majority
of cases, DRM does not prevent data exfiltration and in turn, demonstrates a lot of inconvenience to the consumers while attempting
to provide a theoretical level of protection to content
distributors. As well as this, the extra costs of the development and
maintenance of DRM systems has sometimes meant that paying customers
had to spend more money to ultimately receive a worse functioning
product than a copy of the same product that doesn't use DRM. Although
DRM covers the protection of proprietary data; it does not effectively
protect the needs of an industrial (or enterprise) IoT
deployment. Therefore, an IoT deployment would need more appropriate
technology when it comes to protecting and sharing its sensitive
data. IRM concentrates on these needs and uses end-to-end encryption
to manage individual permissions and usage rights. To this end, Furlong and Cookson proposed methods for managing rights access in a transparent manner~\cite{furlong2006digital}, with a balance between the fraud rate and the insult rate.

Like its predecessor, IRM also has limitations. It was designed to not
only manage access rights to data but also control the interactions
of the consumers of the data to enforce some of these rights. However,
IRM cannot prevent side-channel attacks. For instance, anyone
capturing a photograph of the data once it has been accessed and
displayed on a device. IRM also disallows the use of built-in snapshot
features to prevent digital capture from the same device, but with the
constant advance in technology, however, it is practically impossible
for IRM systems to prevent third-party software from capturing
snapshots. Further, IRM does not prevent data-exfiltration attacks by
malware. If an attacker has stolen the credentials of someone who has
legitimate access to IRM-protected data, an IRM system has no ability
to prevent the data being accessed by the attacker. Aside from
attackers, IRM cannot prevent domain administrators from accessing the
data - which suggests that on an internal level within an organisation
using IRM protection for its data, there must be a higher level of
assurance with technical staff.

\subsection{Availability Attacks on DRM Systems}
Previous work has noted the need for scalability and resilience of
online DRM service
components. Federrath~\cite{federrath2002scientific} and Arnab et
al.~\cite{arnab2004security} note that DRM systems require
authentication with an online rights management server and highlight
the need for scalable server-side DRM
components~\cite{koster2006identity,hauser2003drm}. If this server
suffers from an outage then legitimate consumers of the DRM system
will be unable to access the data protected by the DRM system. Much
more recently, Zhang conducted a survey\cite{zhang2011survey} report
that the use of DRM in particular Ubisoft applications resulted in legitimate users being
denied access due to a severe outage to the Ubisoft DRM. All these
works point to the need for a scalable or global-scale architectures
for DRM deployment that can withstand the challenges of servicing DRM
clients. With the advent of IoT, the challenge becomes even more start
--- Can digital rights management technology offer the super-scaled
architectures required by upcoming IoT deployments?

\section{Architecture}
\label{architecture}

Azure Rights Management (RMS) is a cloud-based service which can be used to
control the flow of data from devices to the cloud, such that
authorised IoT devices and related services can send, receive and
manipulate the data, while others are denied access. Before
data is sent to the cloud, it is encrypted on the device at the
application level, with a policy which defines authorised use of the
data. Some IoT devices may already have several policies pre-packaged
with the device, dependant on the nature of its use. The policy is
typically used to restrict the readability of the data, and restrict
copying and editing only to other devices, organisations and services
stated in the policy. When the protected data is accessed by a
legitimate user, organisation or an authorised service, the data is
decrypted, and the policy attached to the data enforce the rights for
that authorised entity.

\subsection{Protecting Data on an IoT Device}

An RMS client on the IoT device will initially connect to the Azure RMS service which authenticates the device using an Azure Active Directory account. When connected, the authentication is automatic, and the device is not prompted for credentials. After authentication, the connection is then redirected to the organisation's Azure Information Protection tenant. This issues certificates to let the device authenticate to the RMS service, allowing it to protect content offline. The certificates are valid for 31 days provided that the device account is still enabled in the Azure Active Directory.

When protecting data, the RMS client on the device creates a random content key and encrypts the data using the key with AES encryption. The encryption is used for generic protection and native protection when the file is a protected pdf, text or image file (\textit{.ppdf}, \textit{.ptxt} and \textit{.pjpg} respectively). The client then creates a certificate which includes a policy which defines the rights and restrictions of the data, such as an expiration date. The RMS client on the device uses the organisation's key, which was obtained during the initialisation period for the device, to encrypt the policy and the content key. During this time, the RMS client also signs the policy with the certificate already on the device from initialisation. Finally, the client embeds the policy within the encrypted data, allowing the data to be stored and shared anywhere through any means of storage and transmission.

  \begin{figure}[!htbp]
    \begin{center}
   \includegraphics[width=\columnwidth]{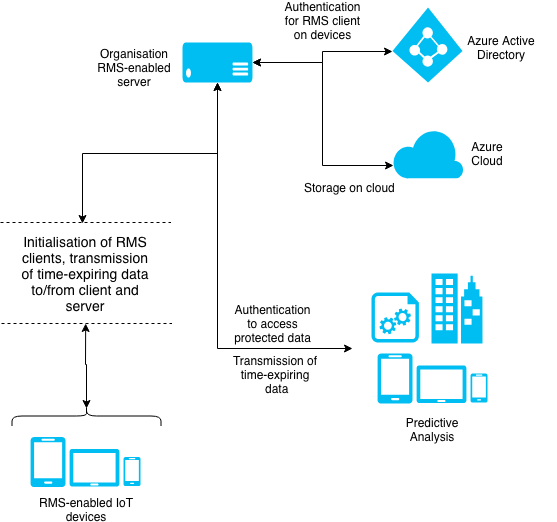}
   \caption{Azure RMS IoT Architecture Diagram~\cite{azurerms}}
    \end{center}
  \end{figure}

\subsection{Accessing RMS-Protected Data}

When an authorised entity, such as those performing predictive
analysis, wants to access the time-expiring data, the RMS client
attached to the service sends the encrypted policy and certificates to
the Azure Rights Management Service. The service decrypts and
evaluates the policy such that a list of rights is obtained specific
to the service. The content key is extracted from the decrypted policy
and is then encrypted with the RMS client's public RSA key obtained
with the request. The content key is then embedded into a use license
with the user rights and is returned to the RMS client. Finally, the
RMS client attached to the predictive analysis service receives the
encrypted use license and decrypts it using its private key, which
also, in turn, decrypts the rights list which is enforced when the
data is accessed. The predictive analysis service can use the data in
many ways, such as running machine and deep learning on the data to
learn and make relevant intelligent decisions.


\section{Denial-of-Service Attacks}
\label{dos}

During initial experiments with Azure's IRM system (i.e. RMS) it was
discovered that changing the clock on the RMS client resulted in it
experiencing stability problems. In particular, when the time on the
client was moved forward by 2 hours the RMS client system crashed and
no applications could view or create IRM protected documents. The only
workable way to recover was to re-install the RMS system on the client
and reboot. We hypothesise that the reason the RMS crashed was related
to the behaviour of such protection measures, and some form of
time-shifting protection would continue to be necessary in future
versions of the Azure RMS client software. To confirm the hypothesis,
we repeated the exercise 25 times. The client crashed on all 25
repetitions giving us confidence that the RMS client crash has a
causal link to the time shift. To confirm with 100\% confidence we
would ideally require access to the client source code, however it is
closed source at the time of writing this paper.

This observation caused us to investigate the reliance of the RMS client
system on synchronized, stable and secure time. It is reasonable to
expect that future versions of client could stop changes in system time
resulting in client becoming unusable. However, since RMS permits
time-limited data, it was clear that there may be more fundamental
dependencies on system time.

The (Azure) RMS client does not need to be connected to the RMS server in
order to read RMS protected data, since the client caches keys
locally and uses these when the server cannot be contacted. If, once
granted, the right to read a document cannot be revoked then this local
caching would not be a problem, however in the case of time-limited
documents the client must be responsible for correctly expiring keys
and re-requesting them from the server. An obvious way of bypassing
the RMS time-limitation restrictions is for a client to
initially open a document within the time period that they are
permitted to have access, so as to obtain a key. Then by manipulating
the local clock, we prevented this key from expiring and so continuing
to have access of the document, even after the time their access
should have been revoked.

In general, these kind of abuses are well known in relation to
time-limited demos of software, and there are a number of ways of
making attacks more difficult. Since a computing device has no
inherent way of maintaining secure time outside of periodic insecure
updates from an external NTP service, these techniques primarily rely
on software watching for unexpected behaviour of the system clock. If
this is detected then access to the protected content is prevented,
although it may be regained if the correct time can then be confirmed
remotely.

Software to detect time-shifting within the OS is typically protected
by code obfuscation and related techniques that apply a graceful
return to a correct clock. While these may be theoretically bypassed,
good techniques exist which would make this task difficult. Our focus
was not so much the myriad ways of continuing to read content after
the associated RMS policy denies this access. Instead we considered
denial-of-service attacks which use the RMS protection which would
prevent users from getting access to protected content, even if the
RMS system should allow them.

These attacks centered on the fact that the computing devices on which
Azure runs uses SNTP (Simple Network Time Protocol) to synchronise
clients to the master clock. Windows used this to synchronize systems
to their domain controller, since secure synchronised time over a
network was required for the use of Kerberos within Active Directory
as an authentication protocol. Windows 8 and Windows 10 extended this
to all machines, by setting systems to synchronize with
\url{time.windows.com} if no domain controller is set.

We observed network traffic between a RMS client and the time
synchronisation server. We found that the optional digital signature
of time update packets from the server was not used. This decision is
understandable since the current standard for authenticated NTP uses
symmetric cryptography and so would allow any machine able to
authenticate time updates to also spoof them. This situation would be
acceptable where all clients trust every other client and the server,
but not the network, however this is not the case with all Windows
devices.

As expected, by spoofing DNS and directing requests for
\url{time.windows.com} to a machine with a SNTP server under our
control allowed us to change the clock on the client machine. However
since the RMS client checks time on a weekly basis it would be necessary to take
control of the DNS server for a long period of time to change the
clock on a significant number of machines. This would be a difficult
task to do on a large scale since eventually any attempt to manipulate
clocks would be noticed, particularly if it resulting in IRM stopping
working.

In order to amplify this attack we investigated ways for this attack
to be performed either on a shorter scale by forcing clients to
update, or without requiring DNS to be taken over. Our initial attempt
was to flood the network with broadcast NTP packets.

\begin{figure}[!htbp]
  \begin{center}
 \includegraphics[width=\columnwidth]{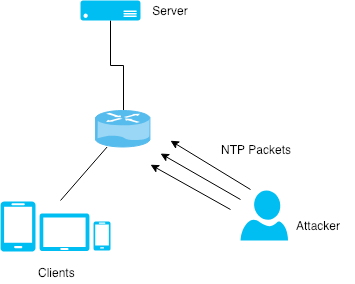}
 \caption{NTP Broadcast Flooding}
  \end{center}
\end{figure}

This did not succeed in changing the time on the clients, we believe for two
reasons. Firstly it seems that a RMS machine only listens for
time updates for a short period after it sends out an update request.
This only occurs every 7 days by default, so machines are only
accepting updates for a tiny proportion of the time. Secondly while
there is no nonce in NTP packets, the request and reply both include
the current time of the client to millisecond precision. The Windows
implementation of the NTP client only seems to accept NTP replies with
the same client time as it included in the request and since the lower
order bits of this are sufficiently unpredictable, a flood approach
unlikely to succeed.

The case of where a RMS machine is part of a domain is slightly
different. According to the Windows Time Service documentation a MAC
key will be negotiated between the domain controller and the client,
and this will be used to sign NTP update packets. In our tests, we saw
NTP packets sent without MACs however it is not clear whether this is
due to a peculiarity of our domain controller.  Additionally we noted
that when the RMS client could not contact the domain controller for
NTP updates, it contacted the DNS server.  Both these observations
point to the fact that the RMS client implementation is not based on
the design principle of secure-by-default.

Since in the domain case, time requests are sent to a machine which
then gets updates from \url{time.windows.com}, one opportunity for
amplification is to put all resources into causing the domain
controller to have the incorrect time.

\begin{figure}[!htbp]
  \begin{center}
 \includegraphics[width=\columnwidth]{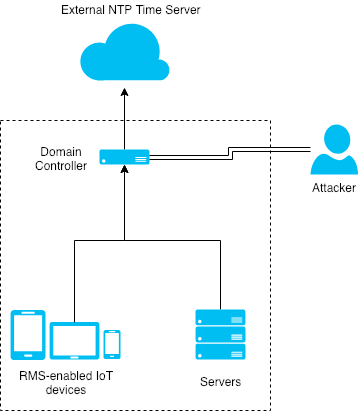}
 \caption{Inducing Incorrect Time on Controller}
  \end{center}
\end{figure}

We found that this propagates to clients,
and so causes the problems with IRM already mentioned.
Furthermore, the Windows Time service documentation states that if a
client detects it is out of sync with the domain controller then it
will update automatically. We found that by changing the domain
controller time, the time on all clients will be changed in a short
time period by the next update cycle. To automate this part of the
attack, we generated spoofed packets in response to NTP packets sent
out by the domain controller purporting to come from the Windows Time
Service (the destination IP being copied from the UDP request
packet). Each spoofed packet carried an offset of four minutes which
was accepted by controller. The default pollrate set on the domain
controller is 4096 seconds, thus requiring just over a day to induce
significant drift.

\paragraph{Defences} The above DoS attacks depend on the attacker being
able to send spoofed SNTP packets to the RMS client. Software-Defined
Networks (SDN) can help mitigate the attack by implementing a simple
threshold-based approach -- reject any NTP response packets reporting
a time offset approaching the Kerberos ticket expiration time. Unlike
conventional switches, an SDN switches can be programmed using an SDN
controller that manages the control plane while the switch focuses on
fast packet-forwarding. The controller runs on a server or a desktop
computing device. SDN developers can write programs that secure and
automate routing logic at the core of the network instead of pushing
this work to firewalls installed at the edge of the network. We
experimented with a programmable PICA8 3290 hardware switch that acted
as the gateway to an RMS client (version 2.1) installed on a desktop
device running Windows 10. We programmed the switch using a Ryu
controller to buffer all NTP packets and redirect them to a switch NTP
proxy-server process listening on port 2100. PICA 3290 is a
Linux-based switch hence it is capable of running Linux binaries,
hence exhibiting intelligence as opposed to a conventional SDN model
where all the intelligence resides on the controller. We wrote socket
code on the switch to drop any NTP packets where the combined values
of NTP offset and NTP delay is more than four minutes. Five minutes is
the threshold beyond which Kerberos rejects client authentication
requests as the tickets expire.

To test the proposed defense, we generated spoofed NTP packets using
the SCAPY tool, purporting to come from the Windows Time Service. We
tested with a range of server offsets between 30 and 600 seconds. The
defense-enabled SDN switch was able to blackhole NTP packets whose
combined offset and delay values were over four minutes with zero
false-positives. Our experiments demonstrate that it is possible to
mitigate attacks using SDN approaches. However, an attacker can
successfully counter our defense by lowering the threshold of change
--- by slowly increasing the NTP offset say at the rate of 30 seconds
at each polling interval. Better statistical approaches might address
this weakness. This will be the subject of future work.

\section{Conclusions}
\label{conclusions}

In this work, we have highlighted the importance of maintaining secure
control of IoT data as it transcends administrative boundaries. We
have examined a popular approach for achieving this, namely via a
cloud-based rights management service from a prominent software
company. We detail several successful service denial attacks via
tampering the time service on which the digital rights management
depends. We report that the attacks are
successful on every attempt without exception. It is particularly
noteworthy that even if IRM system deployed by RMS is made fully
scalable, Denial-of-Service attacks can be mounted by local
adversaries with very little resources. This is a result that has important implications --- when system safety is a function of availability, then a DoS attack on data availability, can  escalate into a safety failure, forcing the operator to engage in an emergency shutdown procedure.


\section{Acknowledgements}
\label{acknowledgements}

The authors gratefully acknowledge funding from EPSRC, UK.


%
%

\bibliography{references}
\footnotesize
\bibliographystyle{abbrv}

\end{document}